\documentclass[superscriptaddress,preprintnumbers,amssymb,twocolumn]{revtex4} 
\pdfoutput=1
\usepackage{graphicx}
\usepackage{epsfig}
\usepackage{amsmath}
\usepackage{amsfonts}
\usepackage{amssymb}

\usepackage{braket}
\usepackage{cancel}
\usepackage{pstricks}
\usepackage{color}
\usepackage{feynmf}

\def\lsim{\:\raisebox{-0.5ex}{$\stackrel{\textstyle<}{\sim}$}\:}
\def\gsim{\:\raisebox{-0.5ex}{$\stackrel{\textstyle>}{\sim}$}\:}

\usepackage{epstopdf}

\begin{document}
{\tiny }\global\long\def\bra#1{\Bra{#1}}
{\tiny }\global\long\def\ket#1{\Ket{#1}}
{\tiny }\global\long\def\set#1{\Set{#1}}
{\tiny }\global\long\def\braket#1{\Braket{#1}}
{\tiny }\global\long\def\norm#1{\left\Vert #1\right\Vert }
{\tiny }\global\long\def\rmto#1#2{\cancelto{#2}{#1}}
{\tiny }\global\long\def\rmpart#1{\cancel{#1}}
{\tiny \par}

\title {A universally enhanced light-quarks Yukawa couplings paradigm}
\author{Shaouly Bar-Shalom}
\email{shaouly@physics.technion.ac.il}
\affiliation{Physics Department, Technion-Institute of Technology, Haifa 32000, Israel}
\author{Amarjit Soni}
\email{adlersoni@gmail.com}
\affiliation{Physics Department, Brookhaven National Laboratory, Upton, NY 11973, USA}

\date{\today}

\begin{abstract}
We propose that natural TeV-scale new physics (NP) with ${\cal O}(1)$
couplings to the standard model (SM) quarks
may lead to a universal enhancement of the Yukawa couplings
of all the light quarks, perhaps to a size
comparable to that of the SM b-quark Yukawa coupling, i.e.,
$y_q \sim {\cal O}(y_b^{SM})$ for $q=u,d,c,s$.
This scenario is described within an effective field theory (EFT)
extension of the SM, for which a potential contribution of certain dimension
six effective operators to the light quarks Yukawa couplings is
$y_q \sim {\cal O} \left( f \frac{v^2}{\Lambda^2} \right)$,
where $v$ is the Higgs
vacuum expectation value (VEV), $v=246$ GeV,
$\Lambda$ is the typical scale of the underlying heavy NP and $f$
is the corresponding Wilson coefficient which depends
on its properties and details.
In particular, we study the case
of $y_q \sim 0.025 \sim y_b^{SM}$, which is the typical
size of the enhanced light-quark Yukawa couplings if the NP
scale is around $\Lambda \sim 1.5$ TeV and the NP couplings
are natural, i.e., $f \sim {\cal O}(1)$.
We also explore this enhanced light quarks Yukawa paradigm in
extensions of the SM which contain TeV-scale
vector-like quarks and we match them to
the specific higher dimensional effective operators in the EFT
description.
We discuss the constraints on this scenario
and the flavor structure of the
underlying NP dynamics and suggest some resulting ``smoking gun" signals
that should be searched for at the LHC,
such as multi-Higgs production $pp \to hh,hhh$ and single
Higgs production in association with a high $p_T$ jet ($j$) or photon
$pp \to hj,h \gamma$ and with a single top-quark $pp \to h t$.
\end{abstract}


\maketitle

\section{Introduction \label{sec1}}

After the discovery of the 125 GeV Higgs-like boson,
one of the main tasks of the current and future runs of the LHC
is to uncover its properties and the physics which underlies its origin.
This has led to considerable effort from both the theoretical and experimental sides,
in the hunt for the NP which may address fundamental
questions in particle physics, 
possibly related to the scalar sector of the SM, such as the observed hierarchy between the
two disparate Planck
and EW scales and the flavor and CP structure in the fermion sector.

The Higgs mechanism of the SM suggests that
the Yukawa couplings of the fermions
are proportional to the ratio between their masses and the Higgs
VEV ($v=246$ GeV), i.e.,
$y_f \propto m_f/v$.
In particular, for the light fermions where $m_f/v$ is vanishingly small,
reactions involving their interaction with the Higgs boson
are in many cases
expected to be strongly suppressed and unobservable in the SM.
Therefore, any observable signal which can be associated with an enhanced
Yukawa coupling of a light fermion would stand out
as clear evidence for NP beyond the SM.
Indeed, current experimental bounds and Higgs measurements do not exclude the
possibility that the Yukawa sector of the SM is modified by TeV-scale
NP that directly affects the couplings of the observed
125 GeV Higgs;
the current bounds do not exclude Yukawa couplings of the Higgs to the light
quarks
of the order of the b-quark Yukawa coupling, i.e.,
allowing $y_{q} \sim {\cal O}(y_b^{SM})$
for $q=d,u,s,c$
\cite{1406.1722,1503.00290,1505.06689,1606.09621,1609.06592,1606.09253}.

In this work we propose a framework where
the Yukawa interactions of all the light quarks are universally enhanced,
naming it the ``Universally Enhanced Higgs Yukawa" paradigm - UEHiggsY paradigm.
In particular, we suggest that, if the pattern and size of the Higgs Yukawa
interaction Lagrangian is controlled by
some TeV-scale underlying NP with natural couplings of ${\cal O}(1)$, then
$y_q \sim {\cal O}(y_b^{SM})$ can be universally realized for all $q=d,u,s,c,b$.
We first describe the UEHiggsY paradigm based on an
EFT approach and then give
an explicit implementation of this mechanism within a renormalizable
prescription involving new TeV-scale vector-like quarks (VLQ) with
natural ${\cal O}(1)$ Yukawa-like couplings to the SM quarks.

\section{An EFT description of the UEHiggsY paradigm \label{UEHiggsYpar}}

Consider the effective Lagrangian piece corresponding
to one of the simplest dimension six effective operators that
can generate non-SM Yukawa-like terms:
\begin{eqnarray}
\Delta {\cal L}_{q H} = \frac{H^{\dagger} H}{\Lambda^2} \cdot
\left( f_{uH} \bar q_L \tilde H u_R + f_{dH} \bar q_L H d_R \right) + h.c.
\label{eq1}~,
\end{eqnarray}
where $H$ ($\tilde H \equiv i \tau_2 H^\star$),
$q_L$ and $u_R,d_R$ are the SU(2) SM Higgs, left-handed quark doublets
and right-handed quark singlets, respectively. Also,
$\Lambda$ is the NP scale and $f_i$ are the corresponding Wilson
coefficients which depend on the details of the underlying NP theory.

When the above dimension six operators are added to the SM Yukawa interaction Lagrangian:
\begin{eqnarray}
{\cal L}_{SM}^{Y} = - Y_u \bar q_L \tilde H u_R - Y_d \bar q_L H d_R + h.c. ~,
\end{eqnarray}
and EW symmetry is spontaneously broken,
one obtains the quark mass matrices $\tilde M_q$ ($q=u,~d$ for up and down-quarks, respectively) and the Yukawa couplings in the weak basis.
The physical quark masses, $M_q$,
are then obtained by unitary rotations of both the left and
right-handed quark fields to the quarks mass basis,
$q_{L,R} \to S^q_{L,R} q_{L,R}$ (the CKM matrix is $V = S^{u\dagger}_{L} S^d_L$):
$M_d \equiv S^{d\dagger}_L \tilde M_d S^d_R = {\rm diag}(m_{d},m_{s},m_{b})$ and
$M_u \equiv S^{u\dagger}_L \tilde M_u S^u_R = {\rm diag}(m_{u},m_{c},m_{t})$, where:
\begin{eqnarray}
M_q = \frac{v}{\sqrt{2}} \left( \hat Y_q - \frac{1}{2} \epsilon \hat f_{qH} \right) ~;~ \epsilon \equiv \frac{v^2}{\Lambda^2}
~,
\end{eqnarray}
and
couplings in the physical quark mass basis are denoted
with a hat: $\hat Y_q \equiv (S^q_L)^\dagger Y_q S^q_R$ and
$\hat f_{qH} \equiv (S^q_L)^\dagger f_{qH} S^q_R$.

The Yukawa couplings, $y_q^{ij} \bar q_i q_j h$, are
then given by:
\begin{eqnarray}
y_{q}^{ij} = \frac{m_q}{v}\delta_{ij} -
\frac{\epsilon}{\sqrt{2}} \left( \hat f_{qH}^{ij} R +
\hat f_{qH}^{ji \star} L \right) \label{yeqmod} ~,
\end{eqnarray}
where $m_q$ is the physical quark mass and $R(L)=(1+(-)\gamma_5)/2$.

It is, therefore, evident from Eq.~\ref{yeqmod} that
our UEHiggsY paradigm is realized if the NP operators in Eq.~\ref{eq1}
are natural, i.e., if $f_{qH} \sim {\cal O}(1)$, and have a typical scale
of $\Lambda \sim {\cal O}(1~{\rm TeV})$. More specifically,
taking $\Lambda \sim 1.3$ TeV and
$\hat f_{qH} \propto f_{qH} \sim {\cal O}(1)$, we
have $\epsilon \hat f_{qH} \sim 0.035$, thus leading to
the UEHiggsY scenario:
\begin{eqnarray}
y_{q} \sim \frac{\epsilon}{\sqrt{2}} \hat f_{qH} \sim 0.025 \sim y_b^{SM} \label{Yq}~,
\end{eqnarray}
for all the light quarks ($q=d,u,s,c$) where
$m_q/v \ll \epsilon \hat f_{qH}$, as well as for the b-quark
for which $m_b/v \sim \epsilon \hat f_{qH}$.$^{[1]}$\footnotetext[1]{Note
that $y_q \sim y_c^{SM}$ would be the natural choice of the UEHiggsY
framework if the NP scale is around 2.5 TeV.}

We note that our UEHiggsY setup which yields the modified Yukawa couplings
of Eq.~\ref{yeqmod}, also allows for a very small
b-quark Yukawa coupling as well as for negative Yukawa couplings
for all light quarks including also
the b-quark. Indeed, a suppressed b-quark Yukawa, e.g.,
of the size of the SM d-quark Yukawa, $y_b \sim y_d^{SM}$,
requires some degree of cancellation between the EFT contribution (with
$\hat f \sim {\cal O}(1)$) and the
SM Yukawa term (with $\hat Y_q \sim {\cal O}(y_b^{SM})$)
to the level of $m_d/m_b$ (see also \cite{1710.00619,1712.07494}). As discussed below,
this fine-tuning is not worse than the typical fine-tuning required
for the UEHiggsY paradigm, e.g., to obtain
$y_d \sim {\cal O}(y_b^{SM})$.
Also, the sign of the Yukawa couplings in the UEHiggsY setup depends
on the sign of the Wilson coefficients, in particular for
the light quarks $q=u,d,c,s$ for which $m_q/v \ll \epsilon \cdot \hat f_{qH}$
when $\Lambda \sim {\cal O}(1)$ TeV and $\hat f_{qH} \sim {\cal O}(1)$.
We note, however, that the dependence of the UEHiggsY signals studied
in section V on the sign of the enhanced $y_q$ is mild,
since interference effects
with the SM are sub-dominant in these processes.

In addition to the modification of the light quarks Yukawa couplings,
the effective operators in Eq.~\ref{eq1} also generate new
tree-level contact interactions between the SM light quarks and two or three Higgs particles, $q \bar q hh$ and $q \bar q hhh$.
These new couplings are also proportional to $\hat f_{qH}$:
\begin{eqnarray}
\Gamma_{\bar q_i q_j hh} =
\frac{3\epsilon}{\sqrt{2}v}  \left( \hat f_{qH}^{ij} R +
\hat f_{qH}^{ji \star} L \right) ~,~
\Gamma_{\bar q_i q_j hhh}  = \frac{\Gamma_{\bar q_i q_j hh}}{v}
\label{yeq2} ~.
\end{eqnarray}
and may cause
large deviations (from the expected SM rates) to the multi-Higgs
production channels $pp \to hh,~hhh$ at the LHC, as will be
discussed in section \ref{sec4}.

The above UEHiggsY paradigm suffers, however, from two potential problems
associated with fine-tuning and flavor:
\begin{description}
\item{\bf{fine-tuning:}} Some degree of fine-tuning is required among the parameters
of the Lagrangian pieces
${\cal L}_{SM}^{Y} + \Delta {\cal L}_{q H}$
in order to simultaneously accommodate
the light-quark masses $m_q \ll m_b$ and
the enhanced Yukawa couplings of $y_q \sim {\cal O}(y_b^{SM})$.
As will be discussed below, this fine-tuning is, however,
not worse than the flavor fine-tuning in the SM.
\item{\bf{flavor:}} The Yukawa couplings $Y_q$ and Wilson coefficients $f_{qH}$ cannot be diagonalized
simultaneously in general. As a result, flavor changing neutral couplings (FCNC) among
the SM quarks may appear. This is manifested
by the off-diagonal elements of $\hat f_{qH}$ (see Eq.~\ref{yeqmod}),
which
are a-priori expected
to be of ${\cal O}(1)$. In particular, with $\Lambda \sim {\cal O}(1)$ TeV, we obtain
FCNC $q_iq_jh$ couplings also of the size of the b-quark Yukawa, e.g.,
$y_q^{ij} \sim \epsilon \hat f_{qH}^{ij}/\sqrt{2} \sim {\cal O}(y_b^{SM})$ for $i=1,j=2$ (see Eq.~\ref{Yq}).
We will address this flavor problem in the next section.
\end{description}

As for the fine-tuning issue, it is typically of the order of $m_q/m_b$, so that the worst
fine-tuning corresponds to the 1st generation quarks, where it is
$\sim {\cal O}(m_{u,d}/m_b) \sim 10^{-3}$. To see that, consider
the mass and Yukawa coupling of a single light quark $q$ in the presence of the interactions terms in ${\cal L}_{SM}^{Y} + \Delta {\cal L}_{q H}$:
\begin{eqnarray}
m_q &=& \frac{v}{\sqrt{2}} \left( Y_q - \frac{1}{2} \epsilon f_{qH} \right) \label{sol1}~, \\
y_q &=& \frac{1}{\sqrt{2}} \left( Y_q - \frac{3}{2} \epsilon f_{qH} \right) \label{sol2}~.
\end{eqnarray}

In particular, fixing $m_q$ to its measured/observed value
(e.g., $m_q \sim 2$ MeV for the u-quark)
and requiring that $y_q \sim y_b^{SM} = \sqrt{2} m_b/v \sim 0.025$,
the solution to Eqs. \ref{sol1} and \ref{sol2} for
the corresponding couplings $Y_q$ and $f_{qH}$ is:
\begin{eqnarray}
Y_q &=& -\frac{y_b^{SM}}{\sqrt{2}} \left( 1- \frac{3}{\sqrt{2}} \frac{m_q}{m_b} \right) \label{choice1} ~, \\
\epsilon f_{qH} &=& -\sqrt{2} y_b^{SM} \left( 1- \frac{1}{\sqrt{2}} \frac{m_q}{m_b} \right) \label{choice2} ~.
\end{eqnarray}
Thus, both $\epsilon f_{qh}$ and $Y_q$ need to be
of ${\cal O}(y_b^{SM})$
and the resulting fine-tuning is at the level of $\Delta_{q} \sim {\cal O}(m_q/m_b)$.
We therefore see that the UEHiggsY paradigm which arises
from natural TeV-scale NP
with ${\cal O}(1)$ couplings, requires
technical fine-tuning of the quark-Higgs interaction
parameters at the level of
$\Delta_{q} \sim {\cal O}(0.1,0.01,0.001)$ for $q=c,s,u/d$, respectively.
In particular, the fine-tuning is at most at the per-mill level and
is only technical in the sense that the fine-tuned parameters,
once fixed, are stable against higher-order corrections (as opposed to the
fine-tuning in the SM Higgs potential).
In fact, this technical $10^{-3} - 10^{-1}$ fine-tuning is comparable to
the flavor fine-tuning problem in the SM, which is manifest
in the CKM matrix that
has no a-priori reason to be close to the identity matrix \cite{1704.06993}.

\section{The underlying heavy physics and flavor \label{sec2}}

\begin{figure}[htb]
\begin{center}
\includegraphics[scale=0.45]{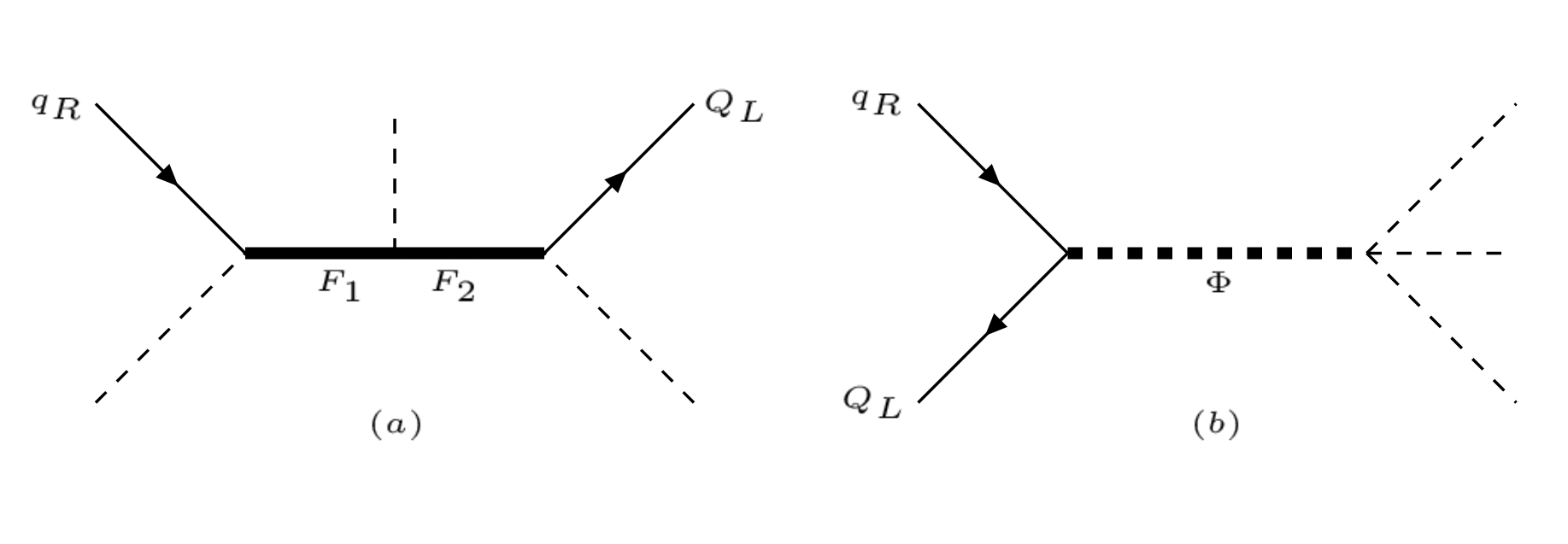}
\end{center}
\caption{Tree-level diagrams in the underlying heavy theory which can generate the dimension 6 operators in Eq.~\ref{eq1}, involving exchanges of heavy VLQ $F_1,F_2$
(left) and a heavy scalar $\Phi$ (right).
See also text.}
\label{diagrams}
\end{figure}

The effective operators in Eq.~\ref{eq1} can be generated by various types of heavy underlying
NP which contain new heavy particles that couple to the SM fermions.
In Fig.~\ref{diagrams} we depict examples of tree-level diagrams
in the underlying theory, which can generate the dimension 6 effective operators of Eq.~\ref{eq1}
when the heavy fields are integrated out. In particular,
the underlaying NP theory may contain
heavy VLQ ($F_1$ and $F_2$) and/or a
heavy scalar ($\Phi$) - both have the required quantum numbers
to couple to the SM quarks and Higgs fields.
Indeed, new heavy scalars and/or vector-like fermions are elementary
building blocks of several well
motivated beyond the SM scenarios which may address
fundamental unresolved theoretical questions in particle physics.

As an example for a simple occurrence of the UEHiggsY
framework,
we will focus below on the heavy VLQ scenario, which
has rich phenomenological implications
\cite{0007316,delaguilar1,dawson1,1304.4219,1306.0572,1404.4398,1406.3349,1703.06134,1710.02325,1712.09360}
and may be linked to the mechanism responsible for solving
the hierarchy problem \cite{EFT_nat}, as well as to naturalness
issues in supersymmetry \cite{VLQ_MSSM} and in
strongly coupled theories where
the light Higgs boson is considered to be a pseudo-Nambu-Goldstone
boson of an underlying
broken global symmetry, e.g., in little Higgs models \cite{VLQ_littlehiggs}
and in models with partial compositeness \cite{1303.5701,1311.2072,1712.07494}.
VLQ dynamics may
also be an important ingredient of the physics
that underlies flavor and CP-violation \cite{1704.06993,delaguilar1,1304.4219,1306.0572,1404.4398,1712.09360,VLQ_flavor}.

In particular, in the VLQ case depicted in diagram (a) of Fig.~\ref{diagrams},
two types of SU(2) VLQ multiplets are required in order
to generate the effective operators of Eq.~\ref{eq1}:
$(F_1,F_2)=({\rm doublet},{\rm singlet})$ and/or $(F_1,F_2)=({\rm doublet},{\rm triplet})$.
We will adopt a SM-like (doublet, singlet) VLQ setup, assuming
three generations of
SU(2) VLQ doublets $Q_i = (U,D)_i$ and the corresponding
up-type and down-type SU(2) singlets $U_i$ and
$D_i$, respectively, carrying
the same quantum numbers as the SM quarks doublets and singlets:
$Q=(3,2,1/6)$, $U=(3,1,2/3)$ and $D=(3,1,-1/3)$.
We assume that the VLQ are in their mass basis, having
explicit mass terms in the full Lagrangian, i.e.,
$M_F (\bar F_L F_R + \bar F_R F_L)$, with
a mass $M_{F=Q,U,D} \sim 1-2$ TeV (the typical
lower bounds on the masses of new VLQ states are
in the range 1-1.5 TeV, depending on their mixing with the SM quarks
and on their decay pattern \cite{VLQbounds}). These
VLQ will also have in general the following Yukawa-like couplings to the SM Higgs (which upon EWSB also give a small contribution to their masses):
\begin{eqnarray}
- {\cal L}_V^Y = \hat\lambda_{QU} \bar Q_L \tilde\phi U_R +
\hat\lambda_{QD} \bar Q_L \phi D_R + h.c. \label{lam1}~,
\end{eqnarray}
where $\hat\lambda_{QU}$ and $\hat\lambda_{QD}$ are $3 \times 3$ matrices in the VLQ flavor space in their mass basis (we have suppressed the generation index of the VLQ).

The Yukawa-like  mixing terms of the VLQ with the SM quarks are
in general:$^{[2]}$\footnotetext[2]{With the VLQ setup presented here the CKM matrix is
extended and the SM $3 \times 3$ CKM block is, in principle, no longer unitary.
However, the deviations from unitarity are expected to be $\propto m_q^2/m_{VLQ}^2$ and, therefore,
very small for $m_q \leq m_b$ and $m_{VLQ} \gsim 1$ TeV, see e.g., \cite{1704.06993}. The details of
such deviations of the SM $3 \times 3$ CKM block from unitarity depend on the
flavor structure of the underlying UV completion which contains the heavy VLQ
fields and is beyond the scope of this work.}
\begin{eqnarray}
- {\cal L}_{Vq}^Y &=& \hat\lambda_{Uq} \bar q_L \tilde\phi U_{R} +
\hat\lambda_{Dq} \bar q_L \phi D_{R} ~, \nonumber \\
&+& \hat\lambda_{Qu} \bar Q_L \tilde\phi u_R +
\hat\lambda_{Qd} \bar Q_L \phi d_R + h.c. \label{lam2}
\end{eqnarray}
where, here also, $\hat\lambda_{Uq,Dq,Qu,Qd}$ are all $3 \times 3$
matrices in the VLQ - SM quarks flavor space and the
SM quark fields are also assumed to be in their physical mass basis.

\begin{table*}[htb]
\begin{center}
\begin{tabular}{|c|}
\multicolumn{1}{|c|}{{\bf \Large $Z_3$ symmetry 1}:
$ \alpha(q_L^k)=\alpha(u_R^k)=\alpha(d_R^k)=(1,2,3)$,
$ \alpha(Q_L^k)=\alpha(D_R^k)=(1,2,0)$,
$ \alpha(U_R^k)=(1,2,1)$ } \\
\hline
$\hat Y_{d},\hat Y_u,\hat \lambda_{QD} \in \left( \begin{array}{ccc}
\times & 0 & 0 \\
0 & \times & 0 \\
0 & 0 & \times
\end{array}\right)$ ~
$\hat \lambda_{QU},\hat \lambda_{Uq} \in \left( \begin{array}{ccc}
\times & 0 & \times \\
0 & \times & 0 \\
0 & 0 & 0
\end{array}\right)$ ~
$\hat \lambda_{Qd},\hat \lambda_{Qu},\hat \lambda_{Dq} \in
\left( \begin{array}{ccc}
\times & 0 & 0 \\
0 & \times & 0 \\
0 & 0 & 0
\end{array}\right)$ \\
$\hat f_{dH},\hat f_{uH} \in \left( \begin{array}{ccc}
\times & 0 & 0 \\
0 & \times & 0 \\
0 & 0 & 0
\end{array}\right)$    \\
\hline
\multicolumn{1}{c}{} \\
\multicolumn{1}{|c|}{{\bf \Large $Z_3$ symmetry 2}:
$ \alpha(q_L^k)=\alpha(u_R^k)=\alpha(d_R^k)=(1,2,3)$,
$ \alpha(Q_L^k)=\alpha(U_R^k)=(1,2,1)$,
$ \alpha(D_R^k)=(1,2,0)$ } \\
\hline
$\hat Y_{d},\hat Y_u \in \left( \begin{array}{ccc}
\times & 0 & 0 \\
0 & \times & 0 \\
0 & 0 & \times
\end{array}\right)$ ~
$\hat \lambda_{QD},\hat \lambda_{Qu},\hat \lambda_{Qd} \in
\left( \begin{array}{ccc}
\times & 0 & 0 \\
0 & \times & 0 \\
\times & 0 & 0
\end{array}\right)$ ~
$\hat \lambda_{QU} \in
\left( \begin{array}{ccc}
\times & 0 & \times \\
0 & \times & 0 \\
\times & 0 & \times
\end{array}\right)$ ~
$\hat \lambda_{Uq} \in
\left( \begin{array}{ccc}
\times & 0 & \times \\
0 & \times & 0 \\
0 & 0 & 0
\end{array}\right)$ ~
$\hat \lambda_{Dq} \in
\left( \begin{array}{ccc}
\times & 0 & 0 \\
0 & \times & 0 \\
0 & 0 & 0
\end{array}\right)$ \\
$\hat f_{dH},~ \hat f_{uH} \in \left( \begin{array}{ccc}
\times & 0 & 0 \\
0 & \times & 0 \\
0 & 0 & 0
\end{array}\right)$    \\
\hline
\multicolumn{1}{c}{} \\
\multicolumn{1}{|c|}{{\bf \Large $Z_3$ symmetry 3}:
$ \alpha(q_L^k)=\alpha(d_R^k)=(1,2,3)$,
$ \alpha(Q_L^k)=\alpha(U_R^k)=\alpha(u_L^k)=(1,2,1)$,
$ \alpha(D_R^k)=(1,2,0)$ } \\
\hline
$\hat Y_{d},\hat\lambda_{Dq} \in \left( \begin{array}{ccc}
\times & 0 & 0 \\
0 & \times & 0 \\
0 & 0 & \times
\end{array}\right)$ ~
$\hat Y_{u},\hat\lambda_{Uq} \in \left( \begin{array}{ccc}
\times & 0 & \times \\
0 & \times & 0 \\
0 & 0 & 0
\end{array}\right)$~
$\hat \lambda_{QD},\hat \lambda_{Qd} \in
\left( \begin{array}{ccc}
\times & 0 & 0 \\
0 & \times & 0 \\
\times & 0 & 0
\end{array}\right)$ ~
$\hat \lambda_{QU},\hat \lambda_{Qu} \in
\left( \begin{array}{ccc}
\times & 0 & \times \\
0 & \times & 0 \\
\times & 0 & \times
\end{array}\right)$ \\
$\hat f_{dH} \in \left( \begin{array}{ccc}
\times & 0 & 0 \\
0 & \times & 0 \\
0 & 0 & 0
\end{array}\right)$ ~
$\hat f_{uH} \in \left( \begin{array}{ccc}
\times & 0 & \times \\
0 & \times & 0 \\
0 & 0 & 0
\end{array}\right)$ \\
\hline
\end{tabular}
\caption{Flavor textures for the fermions Yukawa-like couplings $\hat Y_{u,d},~\hat\lambda_{QU,QD,Qu,Qd,Uq,Dq}$ and the corresponding Wilson coefficients
$\hat f_{uH} = \hat\lambda_{Uq} \hat\lambda_{QU}^\dagger \hat\lambda_{Qu}$ and
$\hat f_{dH} = \hat\lambda_{Dq} \lambda_{QD}^\dagger \hat\lambda_{Qd}$, assuming three different $Z_3$ symmetries due to three types of $Z_3$ charge assignments for the
fermion fields in their mass basis. Our notation for the charge assignments is $\alpha(\psi^k)=(a,b,c)$, using
$k$ as the generation index, so that
$\alpha(\psi^1)=a$, $\alpha(\psi^2)=b$ and $\alpha(\psi^3)=c$.
See also text.}
\label{tab1}
\end{center}
\end{table*}

With this setup, diagram (a) in Fig.~\ref{diagrams} generates
the following $3 \times 3$
Wilson coefficients/matrices $\hat f_{uH},\hat f_{dH}$
(i.e., in the physical quark mass basis)
and effective scales of the operators in Eq.~\ref{eq1}:
\begin{eqnarray}
\hat f_{uH} = \hat\lambda_{Uq} \hat\lambda_{QU}^\dagger \hat\lambda_{Qu} ~&,&~
\Lambda = \sqrt{M_U M_Q} ~, \label{fuh1} \\
\hat f_{dH} = \hat\lambda_{Dq} \hat\lambda_{QD}^\dagger \hat\lambda_{Qd} ~&,&~
\Lambda = \sqrt{M_D M_Q} \label{fdh1} ~.
\end{eqnarray}

Thus, if the VLQ have a mass $M \sim M_U \sim M_D \sim M_Q \sim 1.5$ TeV and natural couplings $\hat\lambda_i \sim {\cal O}(1)$ (so that $\hat f_{qH}^{ij} \sim {\cal O}(1)$), then
the Yukawa couplings
of all light quarks are universally enhanced, with a typical
size of (see Eq.~\ref{Yq}):
\begin{eqnarray}
y_{u}^{ij} &\sim&   \frac{v^2}{M^2}
\left( \hat\lambda_{Uq} \hat\lambda_{QU}^\dagger \hat\lambda_{Qu} \right)^{ij}
\stackrel{\stackrel{{\tiny M \sim 1.5 ~{\rm TeV}}
}{\hat \lambda_{k}^{ij}\sim {\cal O}(1)}}{\longrightarrow} y_b^{SM} ~, \\
y_{d}^{ij} &\sim&   \frac{v^2}{M^2}
\left( \hat\lambda_{Dq} \hat\lambda_{QD}^\dagger \hat\lambda_{Qd} \right)^{ij}
\stackrel{\stackrel{M \sim 1.5 ~{\rm TeV}}{\hat \lambda_{k}^{ij}\sim {\cal O}(1)}}{\longrightarrow} y_b^{SM} ~.
\end{eqnarray}

Therefore, depending on the structure of the VLQ Yukawa-like couplings $\hat\lambda_k$,
potentially ``dangerous" FCNC $q_iq_jh$ transitions of the same size may also be generated, i.e.,
$y_q^{ij} \sim {\cal O}(y_b^{SM})$ for $i \neq j$.

Indeed, FCNC in the down quark sector and among the 1st and 2nd generations of
the up quark sector are
severely constrained by experiment - to the level of
$y_d^{12,21} \lsim 10^{-5}$, $y_d^{13,31,23,32} \lsim 10^{-4}$,
$y_u^{12,21} \lsim 10^{-5}$ \cite{FCNC_constraints}.
This puts stringent constraints on the off-diagonal
elements of the Wilson coefficients $\hat f_{qH}$. In particular,
for $\Lambda \sim {\cal O}(1)$ TeV, these bounds correspond to
$\hat f_{dH}^{ij} \lsim 10^{-3} - 10^{-4}$ for $i \neq j$ and
$\hat f_{uH}^{12,21} \lsim 10^{-4}$, which
therefore constrain the corresponding flavor changing
VLQ coupling to the SM quarks.
This observed smallness of FCNC $q_i \to q_j$ transitions
is a strong indication
that any viable underlying UV completion of the SM, and
in particular of
the above VLQ scenario, should have a mechanism which strongly suppresses or
forbids the above Higgs mediated FC couplings.
Such a mechanism is often assumed to be linked
to an underlying flavor symmetry which gives
flavor selection rules, thus imposing specific flavor
textures on the FCNC couplings.

There are several types of mechanisms and/or flavor symmetries that
can be applied to our VLQ framework, that will give the desired
flavor selection rules. Here we wish to consider
simple and rather minimal examples of flavor symmetries
 which are consistent with both the current experimental constraints on FCNC and with our UEHiggsY framework. In particular, we introduce a $Z_3$ flavor symmetry under which the physical states (i.e., mass eigenstates) of the SM quarks and VLQ fields transform  as $\psi^k \to e^{i \alpha({\psi^k}) \tau_3} \psi^k$, where
$\tau_3 \equiv 2\pi/3$, $k$ is the generation index,
$\psi = q_L,~u_R,~d_R,~Q_L,~U_R,~D_R$ and $\alpha({\psi^k})$
are the $Z_3$ charges of $\psi^k$.

The simplest $Z_3$ setup, which has no tree-level FCNC and also accommodates the UEHiggsY paradigm is the choice $\alpha({\psi^k}) = k$. In this case,
all the Yukawa-like couplings involving the VLQ,  i.e., $\hat\lambda_i$ in Eqs.~\ref{lam1} and \ref{lam2}
as well as the SM Yukawa couplings $\hat Y_{u,d}$ are diagonal,
so that the Wilson coefficients $\hat f_{uH}$
and $\hat f_{dH}$ are also diagonal, giving
$y_q^{ij} \sim y_b^{SM} \delta_{ij}$ for $q=u,d,c,s,b$
and no tree-level FCNC.
In particular, with the $Z_3$ symmetry $\alpha(\psi_k) = k$, the UEHiggsY setup of
Eqs.~\ref{choice1} and \ref{choice2} is realized with only diagonal entries of
$\hat Y_q$ and $\hat f_{qH}$:
\begin{eqnarray}
\hat Y_q^{ii} &=& -\frac{y_b^{SM}}{\sqrt{2}} \left( 1- \frac{3}{\sqrt{2}} \frac{m_{q_i}}{m_b} \right)
\label{choice1_2} ~, \\
\hat f_{qH}^{ii} &=& -\frac{\sqrt{2} y_b^{SM}}{\epsilon}
\left( 1- \frac{1}{\sqrt{2}} \frac{m_{q_i}}{m_b} \right) \label{choice2_2} ~.
\end{eqnarray}
%

In Table \ref{tab1} we list three additional examples of
$Z_3$ symmetries which correspond to
different charge assignments to the fermion fields and yield
non-diagonal structures (textures) for some of the Yukawa-like couplings and Wilson coefficients.
In particular, with the $Z_3$ symmetries 1 and 2
the SM Yukawa couplings $\hat Y_{u,d}$ as well as Wilson coefficients $\hat f_{uH,dH}$ are
diagonal and $f_{uH,dH}^{33}=0$.
Thus, these two flavor symmetries with the
$Y_{u,d}^{11,22}$ and $\hat f_{uH,dH}^{11,22}$ entries of
 Eqs.~\ref{choice1_2} and \ref{choice2_2} and with
$Y_{u}^{33}=\sqrt{2}m_t/v$ and $Y_{d}^{33}=\sqrt{2}m_b/v$,
will bring about the UEHiggsY scenario with
no tree-level FCNC.

The third $Z_3$ symmetry in Table \ref{tab1} generates
a tree-level
$\bar u_L t_R h$ FCNC coupling (due to $\hat f_{uH}^{13} \neq 0$), which is not well
constrained and which may yield an interesting
signal of exclusive production of the Higgs boson in
association with a single top-quark at the LHC. This effect
will be
discussed in more detail in section \ref{subsec4C}.
Notice also that, while the flavor
structures of the SM Yukawa
coupling and Wilson coefficients in the down-quark sector
are similar in all the three $Z_3$ symmetries,
the up-quark sector corresponding to the third
$Z_3$ symmetry has a rank 2 mass matrix,
requiring $\epsilon \hat f_{uH}^{13} = 2 \hat Y_u^{13}$
in order to have a diagonal up-quark mass matrix (i.e., $M_u^{13}=0$).
Thus, in this case there are only two non-zero mass eigenvalues
in the up-quark sector, so that
the UV completion of the VLQ scenario
should have another mechanism for generating the top-quark
mass, e.g., by coupling the top-quark to another
scalar doublet.

\section{Constraints from the 125 GeV Higgs signals \label{sec3}}

The measured signals of the 125 GeV Higgs-like particle are sensitive
to a variety of new physics scenarios, which may alter the Higgs couplings
to the known SM particles involved in its production and decay channels.
In particular,
modifications of the Higgs Yukawa couplings to the light fermions
may lead in general to deviations in both Higgs production and decays.

To see that, we will use the
Higgs ``signal strength" parameters, which are defined as the ratio
between the Higgs production and decay rates and their SM expectations:
\begin{eqnarray}
\mu^{f}_{i} = \frac{\sigma(i \to h \to f)} {\sigma(i \to h\to f)_{SM}} \equiv
\mu_{i} \cdot \mu^{f}
 ~, \label{sig-strength0}
\end{eqnarray}
%
with (in the narrow Higgs width approximation):
\begin{eqnarray}
\mu_i  &=& \frac{\sigma(i \to h)}{\sigma(i \to h)_{SM}} ~,~ \label{Pfac} \\
\mu^f &=& \frac{\Gamma(h \to f)/\Gamma^h}{\Gamma(h \to f)_{SM}/\Gamma^h_{SM}} ~,\label{Dfac}
\end{eqnarray}
where $\Gamma^h(\Gamma^h_{SM})$ are the total
width of the 125 GeV Higgs(SM Higgs),
$i$ represents the parton content in the proton
which is
involved the production mechanism and $f$ is the Higgs
decay final state.

We will consider the signal strength parameters
associated with the production processes
$pp \to h$ and $pp \to h W,~hZ$ followed by the decays
$h \to \gamma \gamma,~W W^\star,~Z Z^\star,~\tau \tau$ and
$h \to b \bar b$, as
analysed by the ATLAS and CMS
collaborations \cite{ss8TeV}.$^{[3]}$\footnotetext[3]{We neglect Higgs production
via $pp \to t \bar t h$, which,
although included in the ATLAS and CMS fits,
are 2-3 orders of magnitudes smaller than the gluon-fusion channel.
Also, the vector-boson fusion (VBF) process $VV \to h$ is not relevant
to our discussion below.}
In the SM, the s-channel production of the 125 GeV Higgs
 is dominated by the gluon-fusion production mechanism $gg \to h$. In particular,
the SM tree-level $q \bar q$-fusion production channel,
$q \bar q \to h$, is negligible due to the vanishingly small
light-quarks SM Yukawa couplings
(the effect of the light quarks in the 1-loop $ggh$ coupling is
also negligible for our purpose, i.e., about $\sim 7\%$ (LO) for the
b-quark \cite{ss8TeV,hgg-vertex,1612.00269,1606.09253}).
In the $pp \to Vh$  channels ($V=W,Z$), the SM rate is dominated
by the s-channel $V$ exchange $q \bar q \to V^\star \to Vh$.

\begin{figure}[htb]
\begin{center}
\includegraphics[scale=0.65]{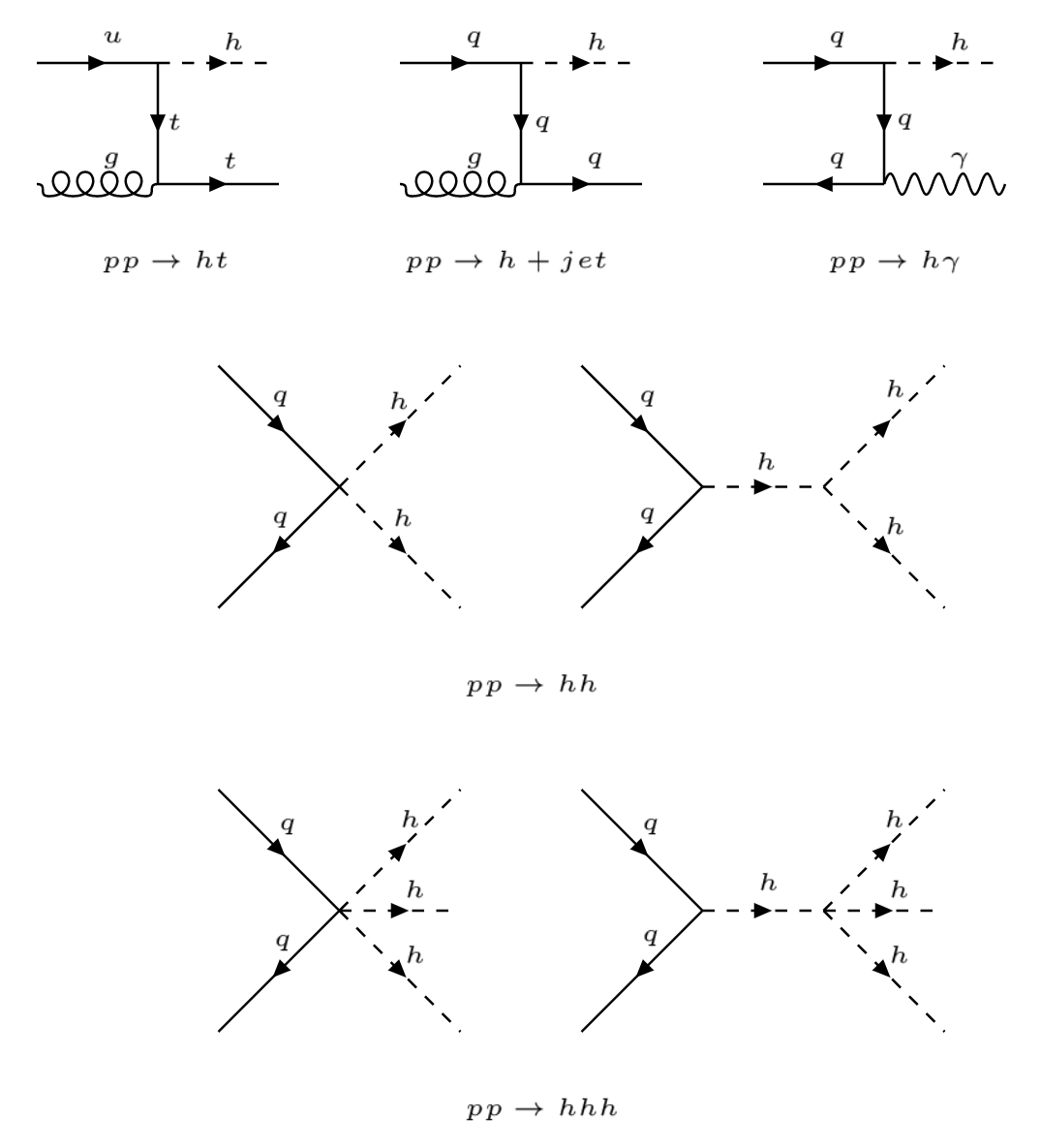}
\end{center}
\caption{Sample diagrams for the processes $pp \to hh,~hhh,~h+jet,ht,~h \gamma$
due to enhanced $qqh$ couplings within the UEHiggsY paradigm.}
\label{NPdiagrams}
\end{figure}

A different picture arises in our UEHiggsY framework, where the Higgs Yukawa couplings
to all the light-quarks ($q=u,d,c,s$) are universally
modified/enhanced.
Higgs production via $q \bar q$-fusion becomes
important, in particular, the tree-level processes $q \bar q \to h$ and
t-channel $Vh$ production
$q \bar q \to Vh$ (see diagram for $q \bar q \to  \gamma h$ in Fig.~\ref{NPdiagrams} and replace
$\gamma \to V$, $V=Z$ or $W$).
To study the effect of these new $q \bar q$-fusion Higgs production channels,
we define Yukawa coupling modifiers, $\kappa_{q}$, and
scale them with the SM b-quark Yukawa, as follows:
\begin{eqnarray}
\kappa_q \equiv \frac{y_q}{y_b^{SM}} ~,
\end{eqnarray}
so that, in the SM, we have
$\kappa_{b}=1$, $\kappa_{c} \sim 0.3$, $\kappa_{s} \sim {\cal O}(10^{-2})$ and
$\kappa_{u,d} \sim {\cal O}(10^{-3})$.
On the other hand, in the UEHiggsY paradigm with a NP scale $\Lambda \sim {\cal O}(1~{\rm TeV})$
and ${\cal O}(1)$ couplings of the heavy states to the SM particles,
we expect $\kappa_q \sim {\cal O}(1)$ for all light-quarks
$q = d,u,s,c$ as well as for the $b$-quark (see discussion below Eq.~\ref{Yq}). In this case
the tree-level $q \bar q \to h$ and $h \to q \bar q$ production and
decay channels
also contribute to
the signal strength factors
$\mu_i$ and $\mu^f$
defined in Eqs.~\ref{Pfac} and \ref{Dfac}.
We neglect below the correction to the 1-loop $gg \to h$ Higgs production channel,
which arises in our UEHiggsY setup from the light-quarks of the 1st and 2nd generations.
As explained below, this correction
is of the order of at most several percent, even
with $y_q \sim y_b^{SM}$ for all $q=u,d,c,s$.
In particular, the contribution of each light-quark
(i.e., in the limit that $m_h^2 \gg m_q^2$) to the 1-loop $ggh$ amplitude
is (see e.g., \cite{0503172}):
\begin{eqnarray}
A_q \propto y_q \cdot \frac{m_q \cdot v }{m_h^2}
\cdot {\rm log}^2 \left( \frac{m_h^2}{m_q^2} \right) ~, \label{Aq}
\end{eqnarray}
and their leading effect to the overall 1-loop gluon-fusion Higgs
production channel arises
from their interference with the top-quark loop (similar
to the case of the leading b-quark contribution in the SM). Thus,
the relative size of any light-quark contribution to the $ggh$ coupling
with respect to that of the b-quark one is:
\begin{eqnarray}
\frac{A_q}{A_b} \sim \frac{y_q}{y_b} \cdot \frac{m_q}{m_b} \cdot
\frac{ {\rm log}^2 \left(\frac{m_h^2}{m_q^2} \right)}
{{\rm log}^2 \left( \frac{m_h^2}{m_b^2} \right) } ~, \label{AqAb}
\end{eqnarray}
so that the contribution to $gg \to h$ from
a $c$($s$)-quark with $y_c(y_s) \sim y_b^{SM}$ is about
50\%(20\%) of the SM b-quark one, i.e., $A_c(A_s) \sim 0.5(0.2) A_b$. Furthermore,
the effect of the light-quarks of the 1st generation is about a hundred times smaller than the
SM b-quark one. Therefore, since the b-quark contribution to the 1-loop $ggh$ production cross-section
is less than $10\%$ (and is included below),
the overall UEHiggsY effect on the $gg \to h$ cross-section is around 5\%
if all the light-quarks have Yukawa couplings $y_q \sim y_b^{SM}$ and is, therefore,
neglected in the analysis below.

Note that, in the decay $h \to \gamma \gamma$, the dominant contribution arises
from the $W$-boson loop and, as a consequence,
the relative effect of the light-quarks loops in our UEHiggsY scenario with $y_q \sim y_b^{SM}$
is much smaller. In particular,
the top-quark loop contributes about 30\%
of $\Gamma(h \to \gamma \gamma)$, mostly
from its interference with the $W$ loop \cite{ss8TeV}. Thus,
for example, the c-quark loop with $y_c \sim y_b^{SM}$ which is
$A_c \sim 0.03 A_t$ (see Eqs.~\ref{Aq} and \ref{AqAb}), will be negligibly small
for our purpose.

In particular, in the UEHiggsY setup we have:
\begin{eqnarray}
\mu_{i=gg+qq}^{UEHiggsY} &\approx&  \frac{\sigma(gg \to h)_{SM} +\hat\sigma(q\bar q \to h)_{UEHiggsY}}{\sigma(gg \to h)_{SM}} \nonumber \\
&\equiv& 1 + \sum_{q} \kappa_{q}^2 R_{q} ~,
\end{eqnarray}
and
\begin{eqnarray}
\mu^f_{UEHiggsY} & \approx & \frac{\kappa_f^2}{1-\left(1 - \kappa_b^2 - \sum_{q} \kappa_{q}^2 \right) BR(h \to b \bar b)_{SM}} ~, \nonumber \\
\label{mudecay}
\end{eqnarray}
where $\kappa_f=g_{hff}/g_{hff}^{SM}$ are the couplings
modifiers of any of the $hff$ Higgs decay vertices
and $R_q$
is defined by the scaled UEHiggsY
$q \bar q \to h$ cross-section evaluated with $\kappa_q=1$
, i.e., using $\sigma(q \bar q \to h)_{UEHiggsY} \equiv  \hat\sigma(q \bar q \to h)_{UEHiggsY}/\kappa_q^2$, as:
\begin{eqnarray}
R_{q} \equiv \frac{\sigma(q \bar q \to h)_{UEHiggsY}}{\sigma(gg \to h)_{SM}} ~,
\end{eqnarray}
where it is understood that $\sigma(q \bar q,gg \to h)$
are convoluted
with the corresponding PDF weights and that $\sigma(q \bar q \to h)_{UEHiggsY}$ are calculated
at tree-level with the values $\kappa_{q}=1$ for all light flavors $q=u,d,c,s$.
Furthermore, in what follows we set the $b$-quark Yukawa coupling to its SM value, i.e., $\kappa_b = 1$,
and neglect the $b \bar b$-fusion production channel
$b \bar b \to h$, which is much smaller than the
light-quark fusion channels, $q \bar q \to h$, when
evaluated with $\kappa_q \sim {\cal O}(1)$.

All cross-sections $\sigma(q \bar q \to h)$ are calculated
using MadGraph5 \cite{madgraph5}
at LO parton-level,
where a dedicated universal FeynRules output (UFO) model
for the UEHiggsY framework was produced for the MadGraph5 sessions
using FeynRules \cite{FRpaper}.
We used the MadGraph5 default PDF set (nn23lo1) and
a dynamical scale choice for the
central value of the factorization ($\mu_F$)
and renormalization ($\mu_R$) scales corresponding to the sum of
the transverse mass in the hard-process.
In particular, we find $\sigma(u \bar u, d \bar d, s \bar s, c \bar c \to h)_{UEHiggsY}
\approx 33.7,23.8,5.4,4.0$ [pb] at the 13 TeV LHC,
so that
using the N3LO QCD prediction (at the 13 TeV LHC) $\sigma(gg \to h) \approx 48.6$ [pb] \cite{ggtoh}, we obtain $\sum_q R_q \sim 1.4$
 and, therefore:
\begin{eqnarray}
\mu_{i=gg+qq}^{UEHiggsY} = 1 + \kappa_q^2 \sum_{q} R_{q} \sim 1+ 1.4 K_{q} \kappa_q^2 ~,
\end{eqnarray}
where we have added a common K-factor, $K_q$, to the tree-level calculated cross-sections
$\sigma(q \bar q \to h)_{UEHiggsY}$.
In particular, with $K_q \sim 1.5$ (see e.g., \cite{1512.04901})
and the UEHiggsY values $\kappa_q=1$ for all $q=u,d,c,s$, we find
that $\mu_{i=gg+qq}^{UEHiggsY} \sim 3$, so that the 125 GeV Higgs production mechanism is enhanced in the UEHiggsY framework
by a factor of ${\cal O}(3)$ with respect
to the SM expectation.

Turning now to the Higgs decay channels
$h \to \gamma\gamma, ~ ZZ^\star, ~WW^\star,~ b \bar b, ~\tau^+ \tau^-$
and assuming no new physics in the decay (by setting $\kappa_f=1$ for $f=\gamma,Z,W,b,\tau$),
we obtain from Eq.~\ref{mudecay}:
\begin{eqnarray}
\mu^{\gamma,Z,W,b,\tau}_{UEHiggsY} =
\frac{1}{1 + 4\kappa_{q}^2 BR(h \to b \bar b)_{SM}} ~. \label{Dparam}
\end{eqnarray}

Thus, under the UEHiggsY paradigm with $\kappa_q=1$
we have $\mu^{\gamma,Z,W,b,\tau}_{UEHiggsY} \sim 0.3$,
so that the calculated signal strengths of
Eq.~\ref{sig-strength0} in these channels are all expected to be the same:
\begin{widetext}
\begin{eqnarray}
\mu^{\gamma,Z,W,b,\tau}_{i=gg+qq}= \mu_{i=gg+qq}^{UEHiggsY} \cdot
\mu^{\gamma,Z,W,b,\tau}_{UEHiggsY}
\approx \frac{1+ 1.4 K_{q} \kappa_q^2}{1 + 4\kappa_{q}^2 BR(h \to b \bar b)_{SM}}
\stackrel{\stackrel{K_q=1.5}{\kappa_q=1}}{\longrightarrow} 0.93 ~. \label{mif_final}
\end{eqnarray}
\end{widetext}

Indeed, the best measured signal strengths
in the four channels
$pp \to h \to \gamma \gamma,~Z Z^\star,~WW^\star,~\tau^+ \tau^-$
have a typical $1 \sigma$ error of 10-20\% and are therefore all consistent with the value
$\mu^{\gamma,Z,W,b,\tau}_{i=gg+qq} \sim 0.93$ within $1-2 \sigma$
(for the LHC RUN1 results see \cite{ss8TeV} and for updated results
from RUN2 see e.g., \cite{meridiani}).
In particular, the currently measured 125 GeV Higgs signals in these four channels do not
constrain the UEHiggsY paradigm with $\kappa_q=1$
for all $q=u,d,s,c$.

Let us next consider the UEHiggsY effect on the measured $hV$ production channel followed
by $h \to b \bar b$. This process has currently the best sensitivity to
the $h \to b \bar b$ decay channel and is used
to overcome the large QCD background
to the simpler $pp \to h \to b \bar b$ channel.
In particular, in this channel we define
$\mu(pp \to hV \to b \bar b V) \equiv R_{hV \to b \bar b V} = R_{hV} \cdot \mu^b$, with ($V=W,Z$):
\begin{eqnarray}
R_{hV} =
\frac{\sigma^{hV}}{\sigma^{hV}_{SM}}
~,
\end{eqnarray}
where $\sigma^{hW},\sigma^{hZ} \equiv \sigma(pp \to hW^+ + hW^-), \sigma(pp \to h Z)$.

As mentioned earlier, in the UEHiggsY framework,
the SM s-channel production process $q \bar q \to V^\star \to hV $ receives additional
tree-level contributions from t-channel $q$-exchange diagrams, similar to the
one depicted for the process $q \bar q \to h \gamma$ in Fig.~\ref{NPdiagrams}.
In particular, calculating the contribution of these
diagrams under the UEHiggsY working assumption with $\kappa_q=1$ for all $q=u,d,c,s$,
we find $R_{hV}^{UEHiggsY} \sim 1.1$ for both $V=W$ and $V=Z$.
Therefore, since
$\mu^{b}_{UEHiggsY} \sim 0.3$ for $\kappa_q=1$ (see Eq.~\ref{Dparam}),
the UEHiggsY signal strength parameter in the
$pp \to Vh \to  b \bar b V$ channel, $R_{hV \to b \bar b V}$,
is expected to be appreciably smaller than one (i.e., than its SM value):
%
\begin{eqnarray}
R_{hV \to b \bar b V} = R_{hV}^{UEHiggsY} \cdot
\mu^{b}_{UEHiggsY} \stackrel{{\kappa_q=1}}{\longrightarrow} 0.33
~, \label{VHres}
\end{eqnarray}
%
for both the $hW$ and $hZ$ production channels.

It is interesting to note that the RUN1 best fitted value for
the measured signal strength in this channel,
$pp \to hV \to b \bar b V$, was indeed on the lower side
and consistent with the above predicted UEHiggsY value
$R_{hV \to b \bar b V} \sim 0.33$
within about $1\sigma$:
the combined ATLAS and CMS analysis of RUN1 data yielded
$R_{hV \to b \bar b V} \sim 0.65 \pm 0.3$ \cite{ss8TeV}.
Recent updated ATLAS and CMS analysis in this channel,
combining the RUN1 data with about 36 fb$^{-1}$ of RUN2
data at a center of mass energy of 13 TeV yielded higher values
$R_{hV \to b \bar b V} \sim 0.9 \pm 0.3$ \cite{1708.03299} and
$R_{hV \to b \bar b V} \sim 1.06 \pm 0.3$ \cite{CMS-PAS-HIG-16-044}, respectively,
but the errors in these channels are still large.

We thus conclude that, currently,
no significant constraints can be imposed on the UEHiggsY paradigm from
the measured 125 GeV Higgs signals. We also note that
the Higgs Yukawa couplings to the light quarks can also effect the
transverse momentum distributions in Higgs production at the LHC
\cite{1606.09621,1606.09253,ourshjpaper}. However, the errors of the
current measured normalized $p_T(h)$ in Higgs + jets production
are still relatively large, so that this analysis
also cannot yet be used to
exclude scenarios with $\kappa_q \sim {\cal O}(1)$
for the light quarks \cite{1606.09621,1606.09253} (see also
discussion in the next section).

\section{Higgs signals of the UEHiggsY paradigm \label{sec4}}

Enhanced light-quark Yukawa couplings
may have direct consequences in Higgs production and decay phenomenology at
the LHC. Indeed, one good example that was discussed in the previous section
is $pp \to Vh$ followed by the Higgs decay $h \to b \bar b$,
which may be sensitive to the UEHiggsY paradigm with improved precision
in the measurement of this Higgs production and decay channel.
Here, we wish to discuss at the exploratory level some of the ``smoking gun"
signals of the UEHiggsY paradigm, associated with the
higher dimension effective operators of Eq.~\ref{eq1}.

Let us define the normalized cross-section ratios:
\begin{eqnarray}
R_{F(h)} \equiv \frac{\sigma(pp \to F(h))}{\sigma(pp \to F(h))_{SM}} ~, \label{RFh}
\end{eqnarray}
where $F(h)$ stands for a final state with at least one Higgs.
In particular, apart from the $pp \to h,hV$ Higgs production channels discussed in the previous section,
the UEHiggsY framework potentially effects other processes
which involve one or more Higgs particles in the final state.
Below we will consider some of the Higgs final states which have a
noticeable tree-level sensitivity to the UEHiggsY paradigm and
which are  also recognized, in general, as sensitive probes
of NP \cite{1610.07922}: Higgs pair and triple
Higgs productions, Higgs  + jets production,
Higgs + single top associated
production and Higgs production with
a single photon, i.e.,
$F(h) = hh,~hhh,~h+nj,~ht,~h \gamma$.$^{[4]}$\footnotetext[4]{Some of the
Higgs signals considered in this section may also be sensitive at 1-loop to modifications
of the 3rd generation Yukawa couplings due to the effective operators in Eq.~\ref{eq1},
see e.g., \cite{1312.3317,1405.7651,1708.00460,1710.00619}.}

Here also, all cross-sections are calculated at LO parton level,
using ${\rm MadGraph5}_{\rm aMC}@{\rm NLO}$ \cite{madgraph5},
with default PDF set and dynamical scale choice for the central value of the
factorization
and renormalization scales.
In addition, following the working assumption of the
UEHiggsY paradigm, the effective operators in Eq.~\ref{eq1} are
assumed to have a typical scale of $\Lambda \sim {\cal O}(1)$ TeV
and couplings $f_{qH} \sim {\cal O}(1)$, so that all cross-sections reported below
are calculated with $qqh$ Yukawa couplings comparable to the SM b-quark Yukawa, i.e., $y_{q} \sim y_b^{SM}$.

\subsection{Multi-Higgs production $pp \to hh,~hhh$ \label{subsec4A}}

Higgs pair production is one of the main targets for NP
searches in the Higgs sector at the LHC, primarily due to its sensitivity to the
Higgs self coupling in the Higgs potential and to heavy NP in the loop induced couplings of
the Higgs to gluons \cite{1406.3349,hh_prod_ref}.
In the SM this process is initiated at LO by 1-loop gluon-fusion diagrams
$gg \to hh$, and the corresponding
cross-section is $\sigma( pp \to hh ) \sim 15$ fb at LO,
where due to the large QCD corrections, it is typically doubled at NLO \cite{NLOhh_calcs}.

In the UEHiggsY framework, there are additional tree-level diagrams
induced by the effective operators of Eq.~\ref{eq1}, as depicted in Fig.~\ref{NPdiagrams}.
Setting $\hat f_{qH}^{ij} =\delta_{ij}$ (i.e., assuming only flavor diagonal couplings)
and $\Lambda \sim {\cal O}(1)$ TeV,
we have $y_q \sim y_b^{SM}$ for the $qqh$ Yukawa coupling
(see Eq.~\ref{Yq}) and
$\Gamma_{hh}^{qq} \sim 3 y_b^{SM}/v$ for the $qqhh$
couplings (see Eq.~\ref{yeq2}).
For this setup we find at LO and for the 13 TeV LHC:
\begin{eqnarray}
R_{hh} \equiv \frac{\sigma(pp \to hh)}{\sigma(pp \to hh)_{SM}} \sim 100 ~, \label{Rhh}
\end{eqnarray}
where more than 90\% of the enhancement arises from the tree-level diagrams initiated
by the u and d quarks. In particular, the total Higgs production cross-section
within the UEHiggsY framework with $y_q \sim y_b^{SM}$ for $q=u,d,c,s,b$
is $\sigma(pp \to hh) \sim 1.5$ pb.

The current best bounds on the $hh$ production cross-section at the 13 TeV
are $R_{hh \to b \bar b \gamma \gamma} \lsim 19$ in the
$hh \to b \bar b \gamma \gamma$ decay channel
(obtained by the CMS collaboration, see \cite{CMS_hh}) and
$R_{hh \to b \bar b b \bar b} \lsim 29$ in the $hh \to b \bar b b \bar b$ decay channel
(obtained by the ATLAS collaboration, see \cite{ATLAS_hh}).

As was shown in the previous section,
in our UEHiggsY framework with $\hat f_{qH}^{ij} =\delta_{ij}$ and $\Lambda \sim {\cal O}(1)$ TeV (for which
$y_q \sim y_b^{SM}$ for $q=u,d,c,s,b$)
the branching ratios for the decays $h \to b \bar b$ and $h \to \gamma \gamma$
are decreased
by about a factor of three
with respect to the SM: $BR(h \to b \bar b , ~ \gamma \gamma) \sim 0.3
BR(h \to b \bar b , ~ \gamma \gamma)_{SM}$ (see Eq.~\ref{Dparam} with $\kappa_q=1$).
Therefore, in these channels we obtain in the UEHiggsY framework:
$R_{hh \to b \bar b b \bar b } = R_{hh \to b \bar b \gamma \gamma} \sim 100 \times (0.3)^2 \sim 10$,
which is an order of magnitude larger than
the SM rate, but still below the current sensitivity.

For the triple Higgs production channel, $pp \to hhh$, the SM cross-section is around
$\sigma(pp \to hhh) \sim 30$ ab at LO and about twice larger at NLO \cite{1610.07922}.
In the UEHiggsY framework (see representative diagrams in Fig.~\ref{NPdiagrams})
we find that $\sigma(pp \to hhh) \sim 10$ [fb], so that:
\begin{eqnarray}
R_{hhh} \equiv \frac{\sigma(pp \to hhh)}{\sigma(pp \to hhh)_{SM}} \sim 300 ~. \label{Rhhh}
\end{eqnarray}

Thus, the expected enhancement over the SM signal
in the $hhh \to b \bar b b \bar b b \bar b$ decay channel
is again $R_{hhh \to b \bar b b \bar b b \bar b} \sim {\cal O}(10)$.
However, since in the UEHiggsY case we have $BR(h \to b \bar b) \sim 0.18$, the
triple Higgs cross-section in this channels is
$\sigma(pp \to hhh \to b \bar b b \bar b b \bar b) \sim 10~{\rm fb} \cdot 0.18^3 \sim 60$ [ab] and,
therefore, might be difficult to detect even at the HL-LHC with a luminosity of 3000 $fb^{-1}$.

\subsection{Higgs + high $p_T$ light-jet production $pp \to h j$ \label{subsec4B0}}

In general, there is a tree-level SM contribution to
the exclusive Higgs + light-jet production, $pp \to h j$,
from the hard processes $gq \to hq$,
$g \bar q \to h \bar q$ and $q \bar q \to hg$, where $q=u,d,c$ or $s$.
However, since the corresponding
tree-level diagrams (see e.g., the t-channel diagram for $gq \to hq$
in Fig.~\ref{NPdiagrams}) are proportional to the light-quarks Yukawa
couplings,
the effect of these light-quark initiated hard-processes on the overall
$pp \to h j$ cross-section is negligibly small in the SM (i.e., when $y_q \ll 1$
in particular for $q=u,d$). Thus,
the dominant SM contribution to the Higgs + light-jet
cross-section arises from the 1-loop gluon-fusion process
$gg \to gh$, which, at leading order, is generated mainly by 1-loop top-quark exchanges.

If, on the other hand, $y_q \sim y_b^{SM}$ for all $q=u,d,c,s$, as
expected in the UEHiggsY framework, then
the contribution (to the $pp \to h j$ cross-section)
from the quark initiated tree-level process $gq \to hq$,
$g \bar q \to h \bar q$ and $q \bar q \to hg$ becomes appreciably larger.
Indeed, in \cite{ourshjpaper} we have shown that the Higgs $p_T$ distribution
in $pp \to hj$ production at the LHC
is a rather sensitive probe of the light-quarks Yukawa couplings
(and also of other forms of NP in the Higgs-gluon $hgg$ and
quark-gluon $qqg$ interactions)
and thus of the UEHiggsY paradigm.

In particular, we have defined in \cite{ourshjpaper} the signal strength
for $pp \to hj$, followed by the Higgs decay $h \to ff$, where
$f$ can be any of the SM Higgs decay
products (e.g.,  $f=b,~\tau,~\gamma,~W,~Z$):
%
\begin{eqnarray}
R_{hj \to f \bar f j} &=& \frac{\hat\sigma(pp \to hj \to f \bar f+j)} {\hat\sigma(pp \to hj \to f \bar f+j)_{SM}} \nonumber \\
&\simeq&
\frac{\hat\sigma(pp \to hj)} {\hat\sigma(pp \to hj)_{SM}} \cdot \frac{BR(h \to f \bar f)} {BR(h \to f \bar f)_{SM}} ~, \label{sig-strength1}
\end{eqnarray}
%
where $\hat\sigma$ is the $p_T$-dependent ``cumulative cross-section", satisfying
a given lower Higgs $p_T$ cut:
\begin{eqnarray}
\hat\sigma \equiv \sigma \left( p_T(h) > p_T^{cut} \right) =
\int_{p_T(h) \geq p_T^{cut}} dp_T \frac{d\sigma}{dp_T} ~, \label{dsigpt}
\end{eqnarray}
and found that, in a NP scenario where $y_q \sim y_b^{SM}$ for all $q=u,d,c,s$
(which corresponds to the UEHiggsY framework discussed here), the above signal strength
is significantly smaller than its SM value at the large $p_T(h)$ regime:
\begin{eqnarray}
R_{hj \to f \bar f j} \sim 0.3 - 0.4 ~,
\end{eqnarray}
for $f = b,~\tau,~\gamma,~W,~Z$ and with a $p_T(h)$ cut in the range
$p_T^{cut} \sim 200 - 1000$ GeV.

\subsection{Higgs-photon associated production $pp \to h \gamma$ \label{subsec4B}}

In the SM, the leading contribution to the exclusive
$pp \to h \gamma$ production channel
is the
tree-level t-channel hard processes $c \bar c,~ b \bar b \to h \gamma$
(shown by the diagram for $q \bar q \to h \gamma$
in Fig.~\ref{NPdiagrams} with $q=c,b$), which give a rather
small cross-section of $\sigma(pp \to h  \gamma) \sim {\cal O}(0.1)$ [fb] with a 30 GeV $p_T(\gamma)$-cut
at the 13 TeV LHC \cite{9706335,1601.03635}. The 1-loop SM (EW) diagrams contributing to the light-quark
annihilation channels, e.g., $u \bar u, d \bar d \to h \gamma$, are more than
an order of magnitude smaller than the tree-level $b \bar b$-fusion
production channel \cite{9706335} and the amplitude for the
gluon-fusion production channel $gg \to h \gamma$ vanishes due to Furry's theorem.

The SM cross-sections for inclusive $h \gamma$ production channels, such as $pp \to h \gamma + j,~ h \gamma + V (V=W,Z), h \gamma + t \bar t,~ h \gamma + t j$ are of ${\cal O}(1)$ [fb] at the 13 TeV, whereas the SM cross-section for the inclusive VBF $h \gamma$ production channel $pp \to h \gamma + 2j$ can reach
$\sim 20$ [fb] \cite{0702119,1601.03635}.

In our UEHiggsY framework, the exclusive channel $pp \to h \gamma$ has an appreciably larger rate due to the
tree-level (t-channel) light-quark fusion diagrams
$q \bar q \to h \gamma$ shown in
Fig.~\ref{NPdiagrams} (i.e., with $q=u,d,s,c$),
which are enhanced by
the ${\cal O}(y_b^{SM})$ $qqh$ Yukawa couplings. In particular, setting
again $\hat f_{qH}^{ij} =\delta_{ij}$ and $\Lambda = 1.5$ TeV
(leading to $y_q \sim y_b^{SM}$), we get
$\sigma( pp \to h \gamma) \sim 1250$ [fb],
at the 13 TeV LHC and with $p_T(\gamma) > 30$ GeV. Thus,
for the exclusive $pp \to h \gamma$ production channel
we find:
\begin{eqnarray}
R_{h \gamma} \equiv \frac{\sigma(pp \to h \gamma)}{\sigma(pp \to h \gamma)_{SM}} \sim 1000 ~, \label{Rhgam}
\end{eqnarray}
where about 80\% of the enhancement arises from the
tree-level $u \bar u$-fusion diagrams.

Here also, taking into account the subsequent
Higgs decay, e.g.,
$h \to b \bar b,~\tau^+ \tau^-,~\gamma \gamma$,
we have $R_{h \gamma \to b \bar b \gamma} =
R_{h \gamma \to \tau^+ \tau^- \gamma} = R_{h \gamma \to \gamma \gamma \gamma} \sim 1000 \times 0.3 \sim 300$,
since the UEHiggsY paradigm only effects
the Higgs Yukawa couplings to the light quarks.

We note that the exclusive $pp \to h \gamma$ channel
is potentially sensitive to other variants of underlying NP
which can be parameterized by different forms
of higher dimensional effective operators,
i.e., other than the ones associated with the UEHiggsY
paradigm in Eq.~\ref{eq1}, \cite{1702.05753}. In particular,
\cite{1702.05753} finds that $\sigma(pp \to h \gamma) \sim {\cal O}(10)$ [fb] can be realized by other types of NP with a typical scale of $\Lambda \sim 1$ TeV and Wilson coefficients of
${\cal O}(1)$. This is more than an order of magnitude smaller than the effect expected in the UEHiggsY case.

Clearly, differential distributions (e.g., such as the photon transverse momentum distribution \cite{1702.05753}) may
provide extra handles for disentangling the various types
of NP that can effect the $h \gamma$ production channel
at the LHC.
This is, however, beyond the scope of this work.

\subsection{Higgs-single top associated production $pp \to t h$ \label{subsec4C}}

The main SM production channels of a Higgs boson in association with a single top quark at hadron colliders are inclusive and have, at LO, two distinguishable
underlying hard processes. These include an extra quark/jet accompanying the $ht$ in the final state \cite{1610.07922}.$^{[5]}$\footnotetext[5]{Another
sub-leading single top production channel in the SM is the
associated production of $th$ with an on-shell W boson in the final state, $pp \to thW$.}
The dominant t-channel process which is initiated by
$bW$-fusion, $b W \to h t + j$, where the extra jet accompanies
the virtual space-like $W$-boson, and the s-channel $q q^\prime$-fusion
hard-process with a virtual time-like $W$-boson, $q q^\prime \to W^\star \to t h +j_b$,
where $q,q^\prime$ are light quarks (i.e., primarily $u,\bar d$ and
$c,\bar s$) and $j_b$ is a b-quark jet.
The t-channel process is very sensitive
to the magnitude and sign of the $tth$ Yukawa coupling \cite{meleth},
and at LO in the SM has a cross-section of
$\sigma(pp \to ht +j)_{SM} \sim 75$ [fb]. The
cross-section for the s-channel process,
$pp \to ht +j_b$, is about 25 times smaller \cite{1610.07922}.

The exclusive $th$ production channels, $pp \to ht$
and $pp \to h \bar t$, involve in the SM
the extremely small 1-loop FC $tuh$ and/or $tch$ vertices
and are, therefore, negligibly small with no observable consequences \cite{EHS1991}.
On the other hand, in the UEHiggsY framework we have for the FC $tuh$
coupling (assuming for simplicity that
$\hat f_{uH}^{13} = \hat f_{uH}^{31}$ ):
\begin{eqnarray}
{\cal L}_{tuh} = \xi_{tu} \bar t u h + h.c.
~~, ~~ \xi_{tu} = \frac{\epsilon}{\sqrt{2}} \hat f_{uH}^{13} ~, \label{xituc}
\label{tuh}
\end{eqnarray}
and similarly for the $tch$ coupling,
where $\epsilon = v^2/\Lambda^2$.
Thus, with $\Lambda \sim 1.5$ TeV and natural underlying
NP (i.e., $\hat f_{uH}^{13} \sim {\cal O}(1)$), we expect the UEHiggsY FC $tuh$
and $tch$ couplings to be typically of
the size of the SM b-quark Yukawa coupling,
$\xi_{tu,tc} \sim y_b^{SM}$, in which case
the exclusive channel $pp \to t h$ has a rate many orders of magnitudes larger
than the SM rate, due to the
tree-level $ug$($cg$)-fusion FC diagrams
$u(c) g \to t h$ (see Fig.~\ref{NPdiagrams}).

\begin{table*}[htb]
\begin{center}
\begin{tabular}{|c||c|c|c|}
  & \multicolumn{3}{|c|}{$\sqrt{s} =13$ TeV (RUN2)} \\
\hline
Higgs signal & SM prediction & our UEHiggsY prediction & Current limit/sensitivity
 \\
\hline \hline
 $R_{hV \to b \bar b V} = \frac{\sigma(pp \to hV \to b \bar b V)}{\sigma(pp \to hV \to b \bar b V)_{SM}}$  & $1$  &
 $\sim 0.33$ & $ \sim 0.9 \pm 0.3$ ({\rm ATLAS} \cite{1708.03299}) \\
 $V=Z,W$ & & & $ \sim 1.06 \pm 0.3$ ({\rm CMS} \cite{CMS-PAS-HIG-16-044}) \\
 \hline
  $R_{hj \to f \bar f j} = \frac{\sigma(pp \to hj \to f \bar f +j)}{\sigma(pp \to hj \to f \bar f +j)_{SM}}$  & & &  \\
  $f=b,\tau,\gamma,Z,W$ & $1$ & $\sim 0.3-0.4$  & None \\
 $p_T(h) >200$ GeV & & & \\
 \hline
$\sigma(pp \to h \gamma)$  & $ \sim 0.1$ [fb] & $\sim 1.25$ [pb] & None \\
$p_T(\gamma) >30$ GeV&  &  & \\
\hline
$\sigma(pp \to h t )$ & $\sim 0$ & $\sim 100$ [fb] &  $\lsim 1.5$ [pb] (CMS \cite{1712.02399})\\
\hline
$R_{hh} = \frac{\sigma(pp \to hh)}{\sigma(pp \to hh)_{SM}}$ & $1$ & $\sim 100$ &  None \\
$R_{hh \to b \bar b \gamma \gamma}$  & $1$ & $\sim 10$ &
   $\lsim 19$ ({\rm CMS} \cite{CMS_hh}) \\
 $R_{hh \to b \bar b b \bar b}$ & $1$ & $\sim 10$ & $ \lsim 29$ ({\rm ATLAS} \cite{ATLAS_hh}) \\
\hline
$R_{hhh} = \frac{\sigma(pp \to hhh)}{\sigma(pp \to hhh)_{SM}}$ & $1$ & $\sim 300$ &
   None \\
$R_{hhh \to b \bar b b \bar b b \bar b}$ & $1$ & $\sim 10$ & None \\
\hline
\end{tabular}
\caption{Some ``smoking gun" Higgs signals of
the UEHiggsY paradigm at the LHC with c.m. energy of 13 TeV.
Also listed are the corresponding SM predictions and
the current limits and sensitivities (from the LHC RUN2)
to some of the signals. The cases where we did not find an experimental
bound/measurement are marked by ``None". The LHC experimental
groups are encouraged to perform a dedicated search in these channels, e.g.,
the exclusive $pp \to h \gamma$, which may also be important for the search of
heavy resonances \cite{hgamma_res}.}
\label{tab2}
\end{center}
\end{table*}

In particular, setting the UEHiggsY values
$\xi_{tu}=\xi_{tc} = y_b^{SM} \sim 0.02$,
we get for the 13 TeV LHC:
$\sigma( pp \to th(\bar t h)) \sim 100(20)$ [fb], with more than 90\%(65\%)
coming from the $ug$-fusion hard-process (i.e., from $\xi_{tu}$).

Defining here the ratios:
\begin{eqnarray}
R_{th/thj} &\equiv& \frac{\sigma(pp \to th)}{\sigma(pp \to th +j)_{SM}} ~, \label{Rth} \\
{\bar R}_{\bar t h/ \bar t hj} &\equiv& \frac{\sigma(pp \to \bar t h)}{\sigma(pp \to \bar t h +j)_{SM}} ~, \label{Rtbarh}
\end{eqnarray}
we find $R_{th/thj}, {\bar R}_{\bar t h/ \bar t hj} \to 0$ in the SM, while
$R_{th/thj} \sim 2$ and ${\bar R}_{\bar t h/ \bar t hj} \sim 0.8$ in the UEHiggsY case.
Notice also that the asymmetric production of $th$ versus $\bar t h$ in the UEHiggsY
framework is different than the corresponding asymmetry in the SM channels $thj$ and $\bar t h j$. In particular,
while in the UEHiggsY case the $th$ production rate is about 5 times larger
than the $\bar t h$ rate, in the SM the $thj$ production rate is less than 2 times larger
than the $\bar t h j$ rate (see \cite{1610.07922}).

Indeed, the CMS collaboration has recently performed a dedicated search for
the exclusive FC single top - Higgs associated production channel $pp \to th$ at
the 13 TeV LHC with a data sample of 35.9 fb$^{-1}$ \cite{1712.02399}. No significant
deviation from the predicted background was observed and bounds on the FC couplings
$\xi_{tu}$ and/or $\xi_{tc}$ were obtained.
In particular, the bounds were reported on the branching ratios
of the corresponding FC decay channels $t \to uh,ch$, which, when translated to the FC couplings
(see derivation below), give $\xi_{tu},\xi_{tc} \lsim 0.09$.
This bound is more than 4 times larger than the expected
strength of these FC couplings in the UEHiggsY framework
 with which the above values
for $R_{th/thj}$ and ${\bar R}_{\bar t h/ \bar t hj}$ were obtained (recall that, within the UEHiggs paradigm, we expect
$\xi_{tu},\xi_{tc} \sim y_b^{SM} \sim 0.02$).
In other words, the current reported sensitivity to the exclusive $th$  final state
is $\sigma(pp \to th + \bar th) \lsim 16 \times \sigma(pp \to th + \bar th)_{UEHiggsY}$,
since the corresponding UEHiggsY predicted cross-section scales as $\xi_{tu,tc}^2$.

Finally, we note that
the current best direct bounds on $\xi_{tu}$ and $\xi_{tc}$
were obtained by the ATLAS collaboration, which analysed the
FC top-quark decays $t \to uh,ch$ in $pp \to t \bar t$ events at
a center of mass energy of 13 TeV and with 36.1 fb$^{-1}$ \cite{1707.01404}. They found
$BR(t \to uh) < 2.4 \cdot 10^{-3}$ and $BR(t \to ch) < 2.2 \cdot 10^{-3}$.

Using Eq.~\ref{xituc},
we have (for $m_{u,c}/m_t \to 0$):
\begin{eqnarray}
BR(t \to uh,ch) \approx \frac{m_t \left( 1-\frac{m_h^2}{m_t^2} \right)}{16 \pi \Gamma_t} \cdot \xi_{tu,tc}^2
\sim 0.57 \xi_{tu,tc}^2 ~,
\end{eqnarray}
where $\Gamma_t$ is the total width of the top-quark.

Thus, the above cited ATLAS bounds translate into the bounds $\xi_{tu},\xi_{tc} \lsim 0.06$,
allowing FC $tuh$
and $tch$ couplings about 3 times larger than the b-quark Yukawa coupling, i.e.,
$\xi_{tu},\xi_{tc} \lsim 3 y_b^{SM}$, which do not rule out the UEHiggsY paradigm with the values
$\xi_{tu},\xi_{tc} \sim y_b^{SM}$.

In Table \ref{tab2} we summarize our predictions for the Higgs signals
considered in this chapter
in the UEHiggsY framework, as well as
the corresponding SM predictions and
the current limits and sensitivities to some of these signals from the LHC RUN2.

\section{Summary \label{sum}}

We have proposed a new framework where the Yukawa couplings of the light quarks of the 1st and 2nd generations, $q=u,d,c,s$, can be as large as the $b$-quark
Yukawa, thus decoupling them from the SM Higgs mechanism, within which a Yukawa coupling of a fermion is proportional to its mass.
We have shown that this scenario (which we named the
``UEHiggsY paradigm") is natural, if the typical scale of the
NP which is responsible for the enhancement of the light quarks Yukawa
couplings is around 1-2 TeV and the heavy (and decoupled) degrees of freedom in the underlying theory have natural couplings of ${\cal O}(1)$ with
the SM quarks.
We have studied the UEHiggsY paradigm in an EFT setup, where dimension
six effective operators yield a Yukawa term
$y_q \sim {\cal O} \left( f \frac{v^2}{\Lambda^2} \right)$, where $\Lambda$ is the typical NP scale and $f$ is a dimensionless coefficient (i.e., the Wilson coefficient in the EFT expansion),
which depends on the properties and details of the underlying
NP dynamics. In particular, with $\Lambda \sim {\cal O}(1)$ TeV
and natural Wilson coefficients $f \sim {\cal O}(1)$, one
obtains $y_q \sim {\cal O}({\rm few}~10^{-2}) \sim {\cal O}(y_b^{SM})$.

We also explore the UEHiggsY scenario in
extensions of the SM which contain TeV-scale
vector-like quarks (VLQ) with a typical mass
of 1-2 TeV, which we matched to
the higher dimensional EFT operators.
We then discuss the flavor structure of the UEHiggsY
Yukawa textures and, in particular, of the
VLQ extension, and the sensitivity of the measured
125 GeV Higgs signals to this paradigm.

Finally, we suggest some ``smoking gun" signals
of the UEHiggsY paradigm
that should be accessible to the future LHC runs:
multi-Higgs production $pp \to hh,~hhh$ and single
Higgs production in association with a high $p_T$ jet or photon
$pp \to hj,h \gamma$ and with a single top-quark $pp \to h t$.

\bigskip
\bigskip

{\bf Acknowledgments:}
We thank Jose Wudka and Arvind Rajaraman for useful
discussions. The work of AS was supported in part by the US DOE 
contract \#DE-SC0012704.

\pagebreak

\end{document}